\documentclass{article}

\usepackage{arxiv}

\usepackage{amsmath,amssymb}

\usepackage[utf8]{inputenc} % allow utf-8 input
\usepackage[T1]{fontenc}    % use 8-bit T1 fonts
\usepackage{hyperref}       % hyperlinks
\usepackage{url}            % simple URL typesetting
\usepackage{booktabs}       % professional-quality tables
\usepackage{amsfonts}       % blackboard math symbols
\usepackage{nicefrac}       % compact symbols for 1/2, etc.
\usepackage{microtype}      % microtypography
\usepackage{lipsum}		% Can be removed after putting your text content
\usepackage{graphicx}
\usepackage{natbib}
\usepackage{doi}

\usepackage{ogonek}
\usepackage{tcolorbox}
\hypersetup{
  pdftitle={Web scraping: a promising tool for geographic data acquisition},
  pdfauthor={Alexander Brenning and Sebastian Henn},
  pdfkeywords={Web scraping, Web mining, Volunteered geographic information, internet data sources, geographic information retrieval},
  hidelinks}
\usepackage{graphicx}

\title{Web scraping: a promising tool for geographic data acquisition}
\author{Alexander Brenning and Sebastian Henn}
\date{Friedrich Schiller University Jena, Department of Geography, and Michael Stifel Center Jena for Data-Driven and Simulation Science (MSCJ), Jena, Germany}

\begin{document}
\maketitle
\begin{abstract}
With much of our lives taking place online, researchers are increasingly turning to information from the World Wide Web to gain insights into geographic patterns and processes. Web scraping as an online data acquisition technique allows us to gather intelligence especially on social and economic actions for which the Web serves as a platform. Specific opportunities relate to near-real-time access to object-level geolocated data, which can be captured in a cost-effective way. The studied geographic phenomena include, but are not limited to, the rental market and associated processes such as gentrification, entrepreneurial ecosystems, or spatial planning processes. Since the information retrieved from the Web is not made available for that purpose, Web scraping faces several unique challenges, several of which relate to location. Ethical and legal issues mainly relate to intellectual property rights, informed consent and (geo-) privacy, and website integrity and contract. These issues also effect the practice of open science. In addition, there are technical and statistical challenges that relate to dependability and incompleteness, data inconsistencies and bias, as well as the limited historical coverage. Geospatial analyses furthermore usually require the automated extraction and subsequent resolution of toponyms or addresses (geoparsing, geocoding). A study on apartment rent in Leipzig, Germany is used to illustrate the use of Web scraping and its challenges. We conclude that geographic researchers should embrace Web scraping as a powerful and affordable digital fieldwork tool while paying special attention to its legal, ethical, and methodological challenges.
\end{abstract}


\renewcommand{\shorttitle}{Geographic Web scraping}

\section{Introduction}\label{introduction}

Since going live in 1991, the World Wide Web (`Web') as part of the broader internet has revolutionized the way in which humans access information. Today, the average internet user spends almost seven hours per day using the internet, or more than 40 percent of their waking time \citep{datareportal.2022}. Web browsers have become a one-stop software for accessing many services, with online platforms increasingly replacing brick-and-mortar businesses and government service desks.

The importance of geospatial data in the Web ---georeferenced data as well as place names in texts--- has continued to gain importance with the advent of the Web 2.0 and the GeoWeb \citep{haklay.et.al.2008.geoweb}. Dynamic Web pages, interactive map displays, location-based services and volunteered geographic information (VGI) are the result of an increased availability of positioning technologies and reflect the importance of location in our lives.

Despite these fundamental and rapid changes in our private and professional lives, research methodologies do not yet fully reflect this radical transformation. In geography and related fields, internet-enabled research strategies mostly relate to conducting Web surveys, leveraging data provided through spatial data infrastructures, or advancing open science by sharing data and code on platforms such as Pangaea \citep{pangaea.2022} for geoscientific data. VGI and crowdsourcing have also emerged as overlapping concepts that describe the increasing availability of user-generated online contents, including geospatial data generated by collaborative mapping initiatives such as OpenStreetMap \citep{goodchild.2007.vgi}. Social media messages from platforms such as Twitter are today also increasingly analyzed with regards to their potential for geographic analyses \citep{albuquerque.et.al.2015}. Unlike the general Web, the platformization of open data sharing and the mining of social media information either strive to establish, or rely on already established standards expressed by contributor guidelines or implemented in the form of application programming interfaces (APIs) faciitating access to Web services.

The broader Web, in contrast, does not know ---or reveal--- such standards, resulting in specific challenges in and techniques for accessing and capturing its contents for subsequent data analysis \citep{glezpena.et.al.2014}. Each website can have a unique structure and follow its own conventions for contents, location references, and structure. This does not make the Web less relevant for academic research: Just like historical vessel logbooks with their semi-structured information can be an invaluable addition to instrumental climate records \citep{schweiger.et.al.2019}, we can learn to extract relevant geographic information from the broader Web in order to complement other means of data acquisition such as surveys or authoritative sources.
Web scraping has therefore emerged as an approach to information retrieval from the Web for a variety of geographic research questions.

Beyond the research context, we would like to point out the potential of Web scraping in teaching geographic information science, especially for experiential learning where students should be exposed to new and unexpected challenges that come with the use of real data. The Web, and therefore Web-enabled data acquisition, offer an ever-replenishing supply of data that can be used to create more variable teaching situations than those offered by textbook datasets.
Moreover, outside the academic context, Web scraping has the potential to provide valuable information for commercial avenues such as geomarketing \citep{boegershausen.et.al.2022} and government applications such as official statistics \citep{hoekstra.et.al.2012}; these share technological approaches and challenges with the academic context, nevertheless their ethical and legal frameworks are partly distinct.

The purpose of this paper is to promote responsible Web-scraping practices as a data acquisition tool for academic research and teaching in geography and related fields. It outlines technological strategies as well as ethical, legal and methodological challenges, and presents a survey of geographic studies involving Web scraping, which may serve as templates for future applications of Web scraping. In contrast to previous overviews in other disciplinary contexts such as psychological, food-price or hospitality research \citep{landers.et.al.2016, hillen.2019.food, han.anderson.2021.overview}, we also highlight aspects that are specific to geographic and geospatial applications, involving implicit and explicit location references and geospatial relationships.

\begin{figure}[!ht]

{\centering \includegraphics[width=14cm]{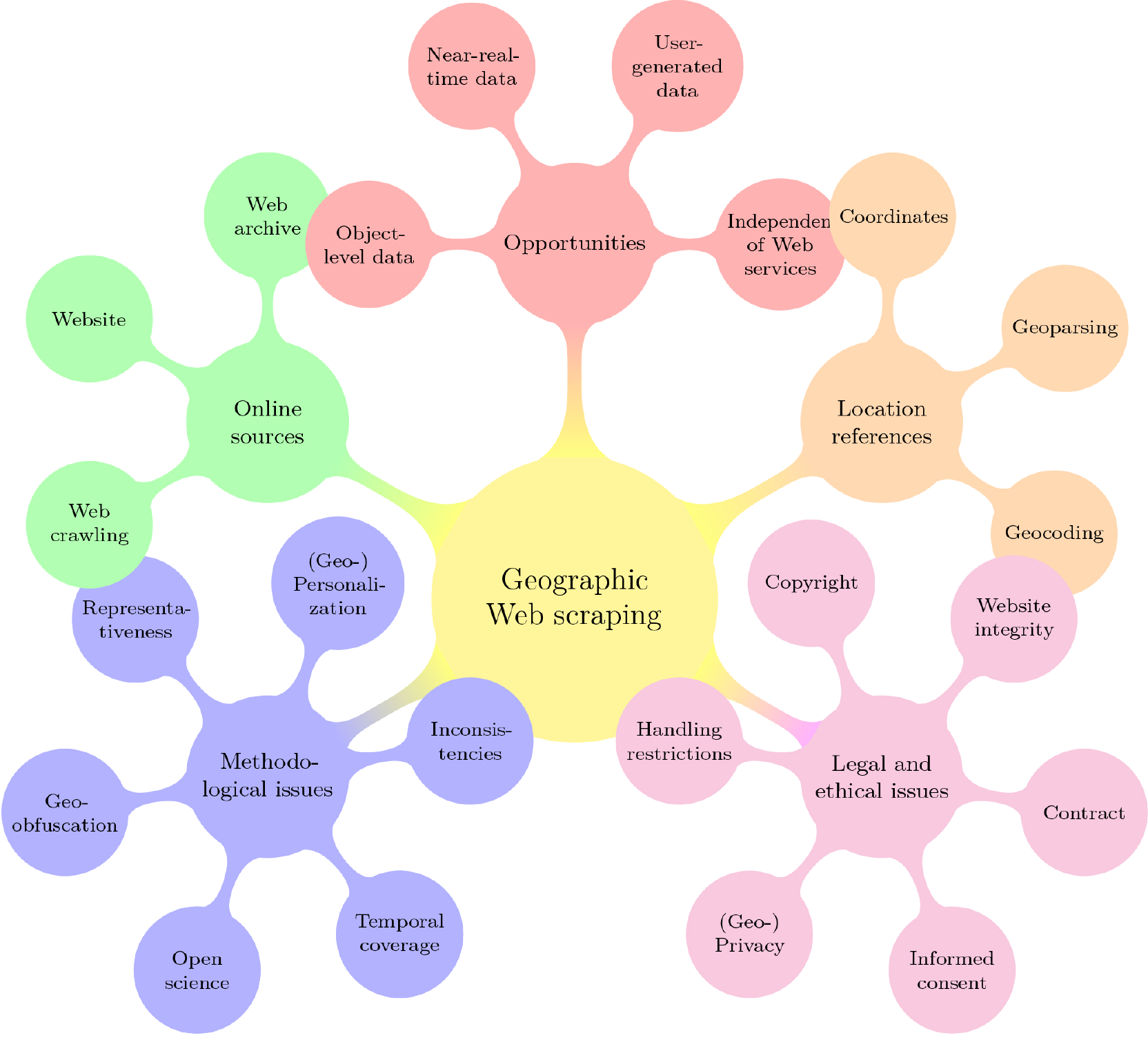}

}

\caption{Mind map showing the various aspects of Web scraping discussed in this paper.}
\label{fig:mindmap}
\end{figure}

This contribution is structured as follows: First, an overview of current opportunities and applications of Web scraping in geography and related disciplines is given.
Based on this, a typical geographic Web scraping workflow and overview of related techniques is presented in order to clarify underlying concepts and technical requirements (Section \ref{workflow}).
The subsequent sections elaborate on legal and ethical issues (Section \ref{legal-ethical}) and methodological challenges affecting data quality, again with a particular focus on spatiality (Section \ref{methodological}). These overarching opportunities and issues are depicted in Figure \ref{fig:mindmap}.
To illustrate the workflow and its challenges in a case study, we will analyze apartment rents from a spatial perspective using Leipzig, Germany as an example (Section \ref{casestudy}), and we finally discuss the potentials and limitations of Web scraping a geographic and geospatial research context.

\section{Opportunities and Application in Geographic Research}\label{opportunities-and-application-in-geographcal-research}

\begin{figure}

{\centering \includegraphics{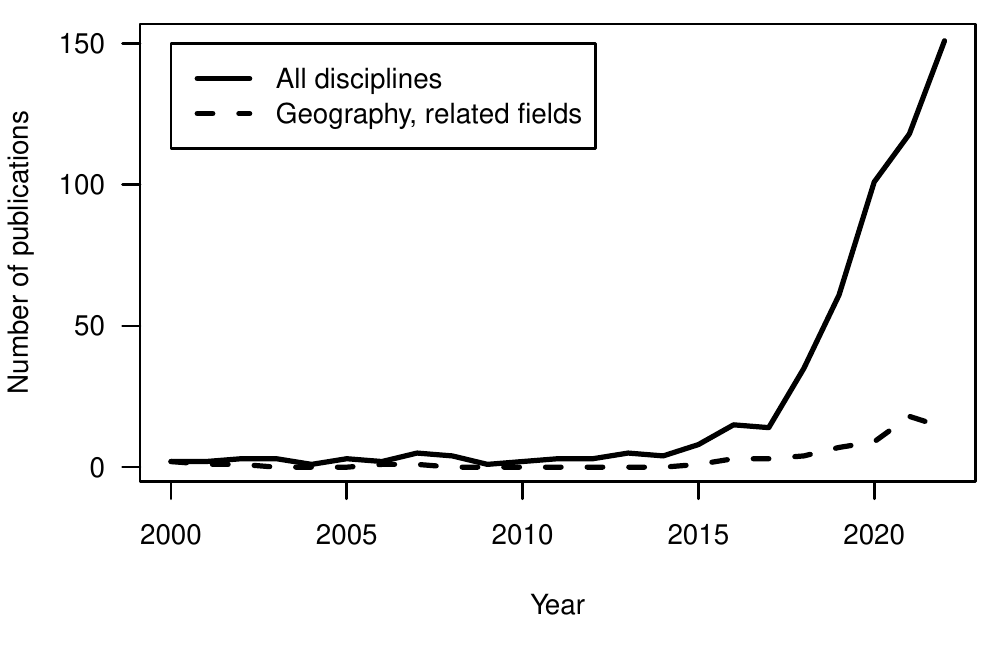}

}

\caption{Adoption of Web scraping in the academic literature since the year 2000. Data source: Clarivative Web of Science Core Collection, (c) Clarivative, 10 May 2023.}\label{fig:wostimeseries}
\end{figure}

Web scraping has only started to gain importance for research in the last five years across all disciplines, including geography and related fields (e.g., planning, tourism, and conservation) according to a search in the Web of Science Core Collection (WoS; search terms `web scraping', `web-scraping' and `webscraping' in `All Fields'; Figure \ref{fig:wostimeseries}). Although it is still a relatively minor phenomenon in these geospatial research areas (65 publications in the WoS) and not all relevant studies are detected based on these search terms, the existing studies allow us to identify research directions and opportunities.

\subsection{Overview of Application Domains}\label{overview-of-application-domains}

The most prominent geography-related application domains include the real-estate market (especially short-term rentals; \citet{han.anderson.2021.overview}) and tourism (Table \ref{tab:immotablereview}). Issues addressed with the help of Web scraping include the analysis of urban transformation in response to a demand for short-term leases \citep{wachsmuth.weisler.2018.airbnb, adamiak.et.al.2019.airbnb, huebscher.et.al.2020}, the classification and mapping of company websites \citep{kinne.axenbeck.2020}, and the spatial analysis of real-estate prices using hedonic models \citep{bonetti.et.al.2016, tomal.2020.cracow, boeing.wadell.2017}, among others. In these fields, Web scraping benefits from the existence of online platforms in markets with a strong platform concentration such as hospitality and real estate.
In physical geography, the focus has mostly been on data acquisition from government websites that do not offer more strongly standardized interfaces such as Web services \citep{bonifacio.et.al.2015, canli.et.al.2018.landslides, samourkasidis.et.al.2019}.

In terms of the type of information scraped, the vast majority of studies focuses on retrieving information on spatial entities (e.g., apartment offers, mosques) and spatial time series (weather data), their characteristics and location references. A relational approach has also been taken to map relationships between companies within an entrepreneurial ecosystem \citep{kinne.axenbeck.2020}, highlighting the potentials of Web scraping ---and of hyperlinks in particular--- to connect various entities.

Beyond academic geography, consumer price research as part of official government statistics has been assessing the potentials of Web scraping for more than a decade in order to automate data retrieval and diversify the data sources \citep{hoekstra.et.al.2012, blaudow.seeger.2019, virgillito.et.al.2019}. Although this is not primarily a geographic research area, it has the potential to be applied at a regional scale to map spatial disparities \citep{benedetti.et.al.2022}. Food-price research is a related area in which the utility of Web scraping was reviewed recently \citep{hillen.2019.food}.

Regarding the parts of the Web that are covered, the surveyed studies focus on information retrieval from the Clear Web, which is open to everyone, and those parts of the Deep Web that are publicly accessible by, for example, dynamically querying information from databases. None of the geographic studies explored the Darknet, which has so far only been scraped for studying online drug markets and cyber crimes \citep[e.g.,][]{crowder.lansiquot.2021.darknet}.

\subsection{Opportunities for Geographic Research}\label{opportunities-for-geographic-research}

Our review of the literature revealed that studies that use scraping techniques benefit from various aspects of the harvested information, creating opportunities for new lines of research in their respective application domains:

\begin{itemize}
\item
  \textbf{Object-level geospatial data.} Since data analyses of individuals or objects at an aggregated level, as offered by most official statistics, may result in ecological fallacies, object-level data is desirable if not necessary for many geographic analyses. This level of detail and the corresponding precision of location information is also essential to obtain relevant micro-geographic attribute information describing, for example, accessibility \citep{tomal.2020.cracow}. Researchers have therefore recognized the potential of Web scraping to provide access to object-level geospatial information by harvesting online platforms in the hospitality or real-estate sectors \citep{wachsmuth.weisler.2018.airbnb, tomal.2020.cracow}, or crawling the Web for company websites and their relations \citep{kinne.axenbeck.2020}.
\item
  \textbf{Near-real-time data.} The scraped online sources often provide data in (near-) real-time, such as the most recent apartment offers or environmental data. Especially socio-economic data such as tourism statistics would otherwise only become available with delay or upon request. Although real-time information is usually not necessary for research (exception: e.g., \citet{canli.et.al.2018.landslides}), this may help to improve the timeliness of research and avoid bottlenecks in data collection.
\item
  \textbf{Access to user-generated content.} User-generated Web contents such as VGI are a (possibly biased) reflection of the interests, motivations and actions of citizens and businesses, but they do not normally become part of public statistics and archives, not even at an aggregated level. Web scraping is the key to these digital media, creating opportunities to study not only user-contributed factual information (e.g., on mosques, {[}gravelle.et.al.2021.mosques{]}), but also the content generators' attitudes \citep{lin.kant.2021} and self-portrayal \citep{schmidt.et.al.2022.greenwashing}.
\item
  \textbf{Independence of Web services.} While governments are striving to implement Web services to share their data, for example geospatial data under the European Commission's INSPIRE directive, Web scraping allows researchers to capture public data that is not (yet) provided in such a standardized form. In other words, the use of Web scraping may be indicative of a lack of such services in areas that are of relevance for geographic research, such as specific types of weather data \citep{canli.et.al.2018.landslides} or public meeting reports \citep{hui.2017.coast}. The same could be said about commercial online platforms, which in some cases provide a fee-based API whose use can be avoided through Web scraping.
\end{itemize}

Overall, while not all of these advantages apply to all use cases of geographic Web scraping equally, and Web scraping may sometimes simply be used due to its cost-effectiveness, they demonstrate that this technique has its place alongside traditional offline as well as online data acquisition methods.

\begin{table}[!ht]
  \centering
  \footnotesize
  \begingroup
    \begin{tabular}{p{2cm}p{3.7cm}p{3.7cm}p{2.5cm}p{2cm}p{2cm}}
      \\
\textbf{Authors} & \textbf{Purpose} & \textbf{Major themes} & \textbf{Scraped information} & \textbf{Scraped websites} & \textbf{Software/Services used} \\
      \toprule
\citet{newing.et.al.2022} & To assess online groceries coverage & Urban--rural inequalities, food deserts & Delivery coverage by postcode & Several online retailers & Python with \texttt{requests} library \\
\citet{wachsmuth.weisler.2018.airbnb} & To assess the extent of Airbnb-induced gentrification & Gentrification trends and the sharing economy & Airbnb via AirDNA & Price, occupancy, location & None \\
\citet{tomal.2020.cracow} & To identify geographic and structural determinants of apartment rent & Non-stationary hedonic modeling & Apartment rent, characteristics, location & \texttt{otodom.pl} & Unknown \\
\citet{kinne.axenbeck.2020} & To map innovation ecosystems & Innovation ecosystems; Web-scraping methodology & Company website text and relations & Crawled company websites & ARGUS based on Python with \texttt{Scrapy} \\
\citet{hui.2017.coast} & To examine use of permitting process in coastal management & Bureaucratic transparency; text mining methodology & Meeting reports incl. task, outcome, address & \texttt{coastal.ca.gov} & Unknown \\
\citet{lin.kant.2021} & To assess the role of social media in citizen participation & Effectiveness of participation in planning processes & Posted messages, their comments, likes, shares & \texttt{facebook.com} & ScrapeStorm \\
\citet{tachibana.et.al.2021} & To reveal the distribution of nature TV programs & Cultural ecosystem services & Broadcasting details, textual summary, extracted toponyms & Archived TV programs at \texttt{nhk.or.jp} & R, \texttt{Rwebdriver}, \texttt{rvest}, Selenium; goo Lab API \\
\citet{schmidt.et.al.2022.greenwashing} & To determine the extent of greenwashing in industry & Air pollution, company self-portrayal & Sustainability-related text fragments & Company websites & Unknown \\
\citet{canli.et.al.2018.landslides} & To interpolate rainfall data in real time & Landslide early warning system & Rainfall measurements at gauging stations & Multiple weather websites & JavaScript with \texttt{MeteorJS}, \texttt{Cheerio}, \texttt{html-parser2} \\
\bottomrule
\end{tabular}
\endgroup
\caption{Overview of selected applications of Web scraping in geographic and related research.}
\label{tab:immotablereview}
\end{table}

\section{The Web-scraping Workflow}\label{workflow}

\begin{figure}[!ht]

{\centering \includegraphics[width=8.5cm]{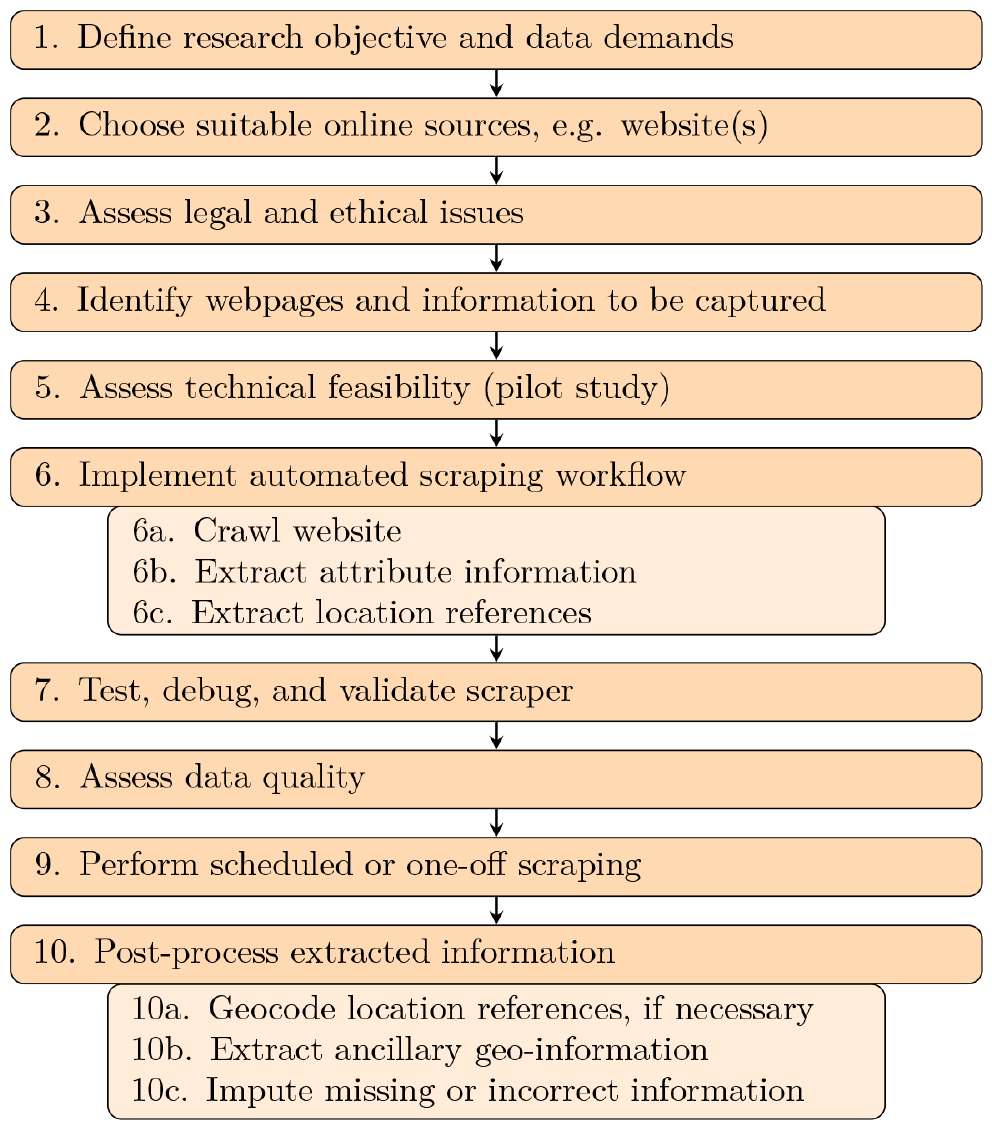}

}

\caption{General outline of the Web-scraping workflow in geographic research. Partly based on \citet{hillen.2019.food}.}
\label{fig:workflow}
\end{figure}

The typical Web-scraping workflow in geographic research, shown in Figure \ref{fig:workflow}, requires a thorough assessment of legal and ethical aspects as well as an evaluation of the technical feasibility of scraping a suitable website, including attention to location references. Considering the challenge of extracting information from websites and webpages whose structure is undocumented to the public and therefore has to be guessed by the researchers, testing, debugging and validation play a critical role. They are often more time-consuming and challenging than in processing datastreams with a well-documented structural and semantic specification. It is, in particular, not uncommon to encounter deviations from the inferred and expected format (or semantic details) after running a scraper for some time.

For geospatial analyses it is of particular importance to extract location references such as place names, complete addresses, or coordinates. Although coordinates are provided in some cases \citep[e.g., ][]{tomal.2020.cracow}, these are usually not displayed verbatim but rather embedded in hyperlinks (URLs) to map displays or in JavaScript data structures that are not actually displayed. Coordinate reference systems are usually not specified, but latitude/longitude information in the World Geodetic System (WGS84) reference are the \emph{de facto} standard (e.g., Google Maps API).

Place names and addresses are useful for obtaining coordinates by toponym resolution or geocoding \citep{melo.martins.2017.geocoding}. When place names are embedded in unstructured text, it may be necessary to use sophisticated algorithms to recognize them beforehand based on gazetteers or machine learning \citep[geoparsing; ][]{hu.et.al.2022.toponym}.

Apart from extracting location references, a relational approach may help to uncover networks relating places and spatial entities to each other. This can be achieved by extracting embedded hyperlinks or entity names. In a recent example, regional company networks representing entrepreneurial ecosystems were reconstructed through crawling and scraping \citep{kinne.axenbeck.2020}. Here, a challenge lies in determining which hyperlinks, or which imprecisely matched entity names, are relevant for the type of relationship of interest (e.g., business partnerships). Machine-learning classifiers can be used for this task \citep{liu.et.al.2020}.

The extraction of other, non-geographic attributes describing the scraped objects is another important task. Established Web-scraping tools are able to extract pieces of information based on structural elements of HTML documents or embedded JavaScript code such as specific tags. Challenges in retrieving numeric data include non-standard and sometimes inconsistent representations of numeric values and their units.

Unstructured text contents, in contrast, need to be represented as a set of features using text mining \citep{munzert.et.al.2014} or topic modeling \citep{thielmann.et.al.2021}. While text mining focuses on extracting features, such as keywords or item frequencies from text, topic modeling can be useful for identifying semantic clusters that may relate to perceptions or discourses referring to places or spatial entities. Sentiment analysis is also of particular interest as it provides a quantitative means to assessing negative or positive emotions that may relate to social characteristics, health outcomes, or marginalization \citep[e.g.,][]{mitchell.et.al.2013.sentiment, delasheras.et.al.2020.sentiment}. Moreover, classification techniques can be used to add attributes to scraped pages or classify them thematically \citep[e.g.,][]{kinne.axenbeck.2020}.

Although Web-scraping software with a graphical user interface (GUI) is increasingly becoming available, programming offers the most flexible approach to retrieving and mining Web contents. Popular environments include the data analysis language R \citep{r.core.team.2021} with its Web-scraping extension \texttt{rvest} \citep{wickham.2021.rvest}. In the Python language, Beautiful Soup \citep{richardson.2022.bsoup} (combined with a HTML parser such as \texttt{HTML.parser}) and the more comprehensive \texttt{Scrapy} library offer similar functionality. Regardless of the chosen programming language, it is important to acquire some basic understanding of HTML, CSS (Cascading Style Sheets), and JavaScript in order to disentangle the structure and meaning of webpages. Proficiency in text processing (e.g., regular expressions and string operations; \citealp{wickham.2019.stringr}) is also essential while more advanced natural language processing and text mining skills \citep{feinerer.et.al.2008.tm, munzert.et.al.2014, han.anderson.2021.overview} are needed for processing unstructured text.

Since many websites require user interaction to load and display relevant contents, software for automated testing of Web applications has also its place in Web scraping, depending on a website's characteristics. Such user interactions may include scrolling a webpage, filling and submitting a Web form, accepting cookies, or logging in as a user. Selenium with its R \citep{harrison.2022.rselenium} and Python libraries is widely used in this context.

Apart from programming, GUI tools available to researchers include mostly commercial visual Web-scraping programs that do not require (traditional) programming skills, such as ScrapeStorm and ParseHub. Typical features include support for interactive websites, cloud storage, database connectivity, IP (internet protocol) address rotation, and VPN (virtual private network) connections. VPN and IP rotation are strategies designed to avoid getting blocked by a server based on scraping patterns or scraper location (see Section \ref{legal-ethical} for possible legal and ethical issues). GUI-based solutions have also been developed in an academic context as open-source software \citep{kinne.2018.argus}.

Secondary providers of Web-scraped data and services have emerged in recent years due to the commercial and academic importance of Web scraping. These include, for example, specialized service providers such as AirDNA \citep{airdna.2022} for monitoring the short-term rental market (Airbnb, Vrbo; e.g., \citealp{wachsmuth.weisler.2018.airbnb}). Coordinated scraping efforts furthermore intend to support communities and increase transparency in the short-term rental market \citep{datahippo.2022, inside.airbnb.2022}.

\section{Ethical and Legal Issues}\label{legal-ethical}

The retrieval and use of third-party data requires the consideration of possible legal as well as ethical issues. These issues primarily relate to how the data is provided and accessed on a website, under what terms of use it is provided, what the economic and privacy implications are, and what consequences the use of the data has for its owner, provider, user, and the objects (or subjects) being described by the data.

\subsection{Legal Aspects}\label{legal-aspects}

Naturally, different rules apply in different jurisdictions, and the physical location of the computer executing the Web-scraping software may be decisive in determining the applicable rules and policies \citep{klawonn.2019}. Recent reviews of legal aspects in the US have been presented by \citet{hirschey.2014}, \citet{macapinlac.2019}, and \citet{stringam.et.al.2021.legal}. \citet{klawonn.2019} has summarized the legal situation in Germany with special emphasis on copyright, and \citet{golla.schoenfeld.2019.recht} with a focus on social media. Other authors provide adopt a broader perspective by providing an overview of legal issues of Web scraping in the `common law world' \citep{liu.2020} or reviewing the regulations in different judicial systems \citep{jevglevskaja.buckley.2023}. Given the complexity of the legal situation and its dependence on the corresponding jurisdiction, we cannot engage in a comprehensive discussion here, but only touch on important issues in three areas:

\begin{itemize}
\item
  \textbf{Copyright.} The legal issues that probably have received the most attention so far relate to copyright law. As a general rule, copyright law requires the consent of an owner before their works may be reproduced by a third party. In principle, the owner of a website does not necessarily own the data it contains, especially when the latter has been generated by users; nevertheless, scraping and republishing the data raise copyright issues time and again \citep{dreyer.stockton.2013}. In this context, it should be noted that it is already the composition of the scraped database itself that may give rise to copyright protection; it may therefore be advisable to use the material in the database according to the principle of `fair use,' i.e., to a limited extent, or to reuse it in a new or original way (e.g., by summarizing it). Further, it may be relevant whether authorization mechanisms, or the lack thereof, constitute a license to copy scraped data. A particularly relevant role accrues to the \texttt{robots.txt} file that provides scrapers with information on possible Web scraping restrictions such as the permitted crawl rate and the areas of the website that are allowed for access \citep{sellars.2018, hillen.2019.food}. Copyright issues of their own kind arise when collecting data from Web archives such as the Wayback Machine \citep{arora.et.al.2016.wayback, nielsen.2016} as their status with respect to copyright is subject to legal challenges (e.g., dyno Nobel Inc v Orica Explosives Technology Pty Ltd (No 2) {[}2019{]} FCA 1552 on 17 September 2019). Webpages from times past may, for instance, not reflect the current legal situation especially in evolving fields such as privacy (e.g., EU General Data Protection Regulation, in effect since 2018). The same is true for the spatial footprint of legality: scraping American Web archives (which may have been scraped using servers located in the US) from outside the US may circumvent an archived website's geoblocking, creating ethical and legal challenges. Researchers should therefore critically consider the feasibility of data sharing \citep[e.g.,][]{datahippo.2022}, and of publishing scraped data as open data (through, e.g., general-purpose platforms such as \citealp{pangaea.2022}).
\item
  \textbf{Contractual compliance.} Another important issue relates to the circumvention of the terms of service (ToS) of a website. This may, e.g., relate to the act of scraping itself, the use of fake user accounts, or of IP rotation services that circumvent technical obstacles such as geoblocking. Whether or not this constitutes a breach of contract is subject to differing interpretations by the courts. It should also be noted that the ToS may not have been perceived or explicitly agreed to by the party performing the Web scraping \citep{dreyer.stockton.2013, zhao.2017}. Some U.S. courts therefore do not consider a mere circumvention of the ToS to be a criminal offense under the Computer Fraud and Abuse Act (CFAA) designed to fight computer crime and hacking \citep{macapinlac.2019} if the user in question has not taken an affirmative action on the website to become a party under the applicable ToS \citep{krotov.et.al.2020.legal}.
\item
  \textbf{Website integrity.} If repeated access to websites results in the service in question being interrupted, liability issues may arise \citep{hirschey.2014, zhao.2017}. Specific legal and also ethical \citep{han.anderson.2021.overview} issues arise in this context when other users are prevented from accessing the website as a result of such activities. In general, as server capacities and bandwidths have developed, this problem has recently become less important \citep{thelwall.stuart.2006.ethical}.
\end{itemize}

On the whole, as Web scraping is a comparatively new research technique, a regulatory framework is still evolving \citep{hillen.2019.food, han.anderson.2021.overview}, and the case law to date is often inconsistent \citep{krotov.et.al.2020.legal, brewer.et.al.2021}. In this context, recommendations, for example those developed by European statistics offices \citep{condron.et.al.2019}, may prove to be a useful resource for researchers. In any case researchers should protect themselves from potentially disadvantageous legal consequences and take appropriate precautions before carrying out Web scraping activities \citep{landers.et.al.2016, hillen.2019.food}, also in the context of addressing geographic research questions.

\begin{figure}
\begin{tcolorbox}
\begin{enumerate}
\item Are there viable alternatives to scraping the data from a website?
\item Do the website's terms of service explicitly prohibit Web scraping?
\item Does the website identify the copyright holder and define a license under which the contents are provided?
\item Can scraping potentially cause material damage to the website or the Web server that hosts it?
\item Has the website blocked or restricted the user's access to its contents or asked the user to cease and desist?
\item Does the website's \texttt{robots.txt} protocol limit or prevent Web-scraping activities?
\item Is the scraped data only a small fraction of the website's database contents?
\item Can the data obtained from the website compromise individual privacy, research subjects' rights, or non-discrimination principles?
\item Can the scraped data reveal confidential information about organizations affiliated with the website?
\item Can the project that requires the Web data potentially diminish the value of the service that the website provides?
\item Does the quality of the data obtained from the Web have the potential to lead to ill-informed decision making?
\end{enumerate}
\end{tcolorbox}
\caption{Questions for the assessment of the legal and ethical viability of Web-scraping projects. Modified and expanded after \citet{ krotov.et.al.2020.legal}, (c) Association for Information Systems.}
\label{fig:legalethical}
\end{figure}

\subsection{Ethical Aspects}\label{ethical-aspects}

Even Web scraping activities that are legally unproblematic may raise a number of ethical questions. These have been reflected in the academic discourse \citep[e.g.,][]{landers.et.al.2016, hand.2018, brewer.et.al.2021, stringam.et.al.2021.legal} but also been the subject of various guidelines that can be found online \citep[e.g.,][]{suciu.2021, thakur.2022}. Important ethical issues related to Web scraping refer, amongst other things, to the following aspects:

\begin{itemize}
\item
  \textbf{Informed Consent.} From an ethical point of view, classical offline research requires implicit or explicit consent from the ones being researched \citep{miggelbrink.et.al.2022}. For most Web scraping activities, however, this consent does not exist. This does not necessarily have to be a problem since for ethical reasons, consent can be dispensed with if the expected benefits of the research exceed the risks associated with it \citep{brewer.et.al.2021}. However, it may be very difficult to determine this in individual cases, for example because it is not possible to identify precisely which information that appears online can reasonably be classified as ``private'' and which as ``public.'' In other words, even if certain text passages or other media can be found online, the content could be private or contain information that, on its own or in combination with other data, such as location, could even allow drawing conclusions about individuals \citep{mahmud.et.al.2014}. One important question that arises in the context of informed consent is therefore whether information accessible online has the character of public information. It is also worth questioning to what extent it can be considered ethical to be part of a forum for the sole purpose of collecting certain data without participating in the conversations that take place there, as this establishes power asymmetries while at the same time blurring the line between private and public information \citep{sugiura.et.al.2017}.
\item
  \textbf{Privacy.} An essential question is whether the data collected allow conclusions to be drawn about individual persons. Following the basic ethical principle of avoiding harm \citep{brewer.et.al.2021, miggelbrink.et.al.2022} can thus require the removal or the modification of all identifiers from the data that could be associated with individuals. These can be, for example, names, hyperlinks but also IP addresses and mobile device identifiers \citep{sugiura.et.al.2017, hand.2018}. Often, however, this is not sufficient, as verbatim text passages can be easily found through internet searches and associated with the persons who wrote them. Therefore additional, in some cases time-consuming processing steps may be required. This may involve summarizing the original data, or extracting relevant features through text mining. This whole issue is further complicated by the fact that there are people who do not want to be anonymized (e.g., bloggers). In such cases any removal of identifiers could thus constitute a copyright infringement \citep{sugiura.et.al.2017}.
\item
  \textbf{Handling access restrictions.} From an ethical point of view it may be acceptable to disregard an explicit prohibition to collect data stated in the ToS if the benefits associated with web scraping are found to outweigh potential harms. Specifically, it may be argued in some cases that there is no other, or no similarly cost-effective way to scientifically analyze a given subject matter. Similar considerations apply to ignoring robots exclusion protocols in the \texttt{robots.txt} file \citep{brewer.et.al.2021}. However, it is important to keep in mind that activities that violate ToS may constitute a criminal offense even if they have been determined to be ethical \citep{brewer.et.al.2021}.
\end{itemize}

To ensure that their research is ethically sound, researchers should not carry out Web scraping at random, but only after careful consideration of the expected benefits and potential risks associated with their activities. This requires a thorough examination of the issue of data collection and processing, as well as a preference for presenting data at an aggregate rather than individual level. In this context, it should also be noted that researchers themselves may be ethically affected by their own research activities. Especially the Darknet with its high degree of anonymity is a place in which geographic and social-science research can observe a range of otherwise hidden, marginalized or possibly criminal activities \citep{benjamin.et.al.2019}. Being exposed to evidence of unethical or illegal activities may lead to traumatization \citep{brewer.et.al.2021}, and Web scraping (and Web crawling) as an automated activity offers little protection against entering problematic zones within the Darknet.

A good exercise and starting point is to assess the legal and ethical viability of a Web-scraping activity based on a set of questions. \citet{krotov.et.al.2020.legal} presented such a set of questions, which we have slightly modified and enhanced (Figure \ref{fig:legalethical}). Negative answers to any of these questions should be a warning sign that should encourage researchers to critically assess the viability of a project, perhaps consulting legal or ethical experts. Nevertheless, this does not automatically mean that the Web-scraping project in question is per se illegal or ethically problematic.

\section{Methodological Challenges}\label{methodological}

Web scraping of geospatial data faces a combination of challenges, some of which are due to the scraping process itself while others are inherent in the data sources being harvested. In our overview we especially highlight issues that are related to the Web scraping process and/or to location and spatial patterns, although these aspects are partly intertwined with biases related to the data sources themselves. The literature on data quality of VGI from social media also offers relevant insights that are more broadly related to capturing geospatial online data \citep[e.g.,][]{tjaden.2021}. Important issues include the following:

\begin{itemize}
\item
  \textbf{Limited dependability.} The structure and contents of websites may change at any time without prior notice, and service providers may impose technological obstacles at any time. Researchers must therefore be able to rapidly adjust scraping software, which may imply significant software development efforts. Also, legal and ethical aspects may have to be re-evaluated regularly. As an example, the German real-estate platform ImmoScout24 blocked scraping activities on its website in 2020 while establishing a fee-based API. Authors also reported software maintenance in response to changes in website structure in Airbnb (2018/19) or targeted climate data portals \citep{bonifacio.et.al.2015, slee.2020}.
\item
  \textbf{Incompleteness.} Scraped information may present significant gaps either due to technical challenges in scraping the available information, or due to incomplete data being provided on a website. In particular, it is important to recognize that attributes that are important to the scraper may be less relevant to the data provider. As an example, real-estate agents may have an interest in omitting apartment characteristics that may lower a property's value. Information may also be entered without proper standardization or validation, e.g.~with rare and unexpected qualifiers (`approx.', `at least'), or non-standard address information involving informal or abbreviated toponyms. As a result, scrapers must be robust to unexpected formats, detect anomalies and outliers, and perform sanity checks. In the example of \citet{bonetti.et.al.2016}, 29\% of the scraped real-estate offers could not be geocoded, and in the end only 28\% of the records were left due to missing attribute values or outliers. Gap filling (imputation; \citealp{gelman.hill.2006}) is often a necessary step that must be used with great care in order to balance robustness against the need to detect systematic problems.
\item
  \textbf{Obfuscation of location.} The accuracy of geographic location is sometimes intentionally degraded to protect location privacy or business interests. This can be achieved by adding a random positional error (e.g., Airbnb: 150--200 m; \citealp{deboosere.et.al.2019}), reporting the location of higher-level spatial units, or providing incomplete address information (e.g., house number unavailable in 20\% of the apartments scraped in our case study in Section \ref{casestudy}). Obfuscation affects, in particular, the calculation of micro-geographic descriptors for spatial modeling, such as walkability. Location effects may thus be underestimated due to regression dilution \citep{frost.thompson.2000.correcting}. It is, however, sometimes possible to infer the true location if obfuscation is poorly implemented \citep{ardagna.et.al.2011.obfuscation}, or to reduce its effects on an analysis with application-specific correction and aggregation strategies \citep{wachsmuth.weisler.2018.airbnb}.
\item
  \textbf{Search personalization and geotargeting.} Web sites use various sources of information to personalize their contents and search results \citep{micarelli.et.al.2007, teevan.et.al.2010.personalization}. This may involve spatial (location of device: geotargeting, geoblocking), temporal (time of search), and thematic context information that informs models of user needs. It may influence the order of search results, their accuracy and completeness, and even the attributes themselves (e.g., personalized and geographic pricing). Web scrapers can to some extent mitigate these effects by emulating a variety of browsers, varying the time of search, rotating IP addresses, using virtual private network (VPN) services to bypass geoblocking, and creating fake user profiles. However, some of these strategies raise ethical and legal questions.
\item
  \textbf{Representativeness.} Web-scraped data often suffers from various biases related to the representativeness of the data. This may relate to the scraping process itself in unexpected ways as, for example, the Airbnb scraper of \citet{slee.2020} appeared to have missed some of the listings in high-density areas. As far as the data sources themselves are concerned, data obtained from online marketplaces does not necessarily represent the broader market due to undercoverage \citep{beresewicz.2017}, the severity of which may vary spatially depending on a platform's market penetration. Offers that remain available for a longer period of time will also be over-represented in one-off data capture. Achieving consistency with other independent data sources is therefore challenging; this is especially relevant for official government statistics or when combining various data sources \citep{agarwal.et.al.2019}.
\item
  \textbf{Logical inconsistencies.} Web scraping over extended periods of time or in platforms covering large regions runs the risk of producing inconsistent data. A platform's internal data collection and validation procedures may change at any time and are usually undocumented. Thus, the publicly accessible attributes may change in quality, semantics, or even availability at any time. As an example, Airbnb reservation status could be scraped until late 2015; in contrast, occupancy in AirDNA's more recent Airbnb data is estimated using machine learning \citep{deboosere.et.al.2019}. These as well as other changes \citep{alsudais.2021} may be more subtle and harder to detect than changes in a website's fundamental structure.
\item
  \textbf{Limited temporal coverage.} Instead of scraping a platform continuously over an extended period of time in a longitudinal design, researchers would often like to access historical data retrospectively. Web archives such as the Internet Archive's Wayback Machine offer access to numerous snapshots of webpages and online media. Going as far back as the year 2001, the Wayback Machine has been identified as a useful resource for the social sciences \citep{arora.et.al.2016.wayback} that has also been leveraged in geographic research contexts \citep[e.g.,][]{tachibana.et.al.2021}. Nevertheless, Web archives are a substantially incomplete copy of the scrapable portions of the Web as they usually do not cover the Deep Web, such as dynamic content and scripted content that is generated in response to user interaction or queries \citep{arora.et.al.2016.wayback}. Dynamic webpages are, however, particularly important as interfaces to large databases such as real-estate listings. Also, unique legal and ethical issues may arise as a consequence of archiving activities (see Section \ref{legal-ethical}).
\item
  \textbf{Barriers to open science.} Depending on the legal space in which Web scraping takes place, researchers may not be able to share the retrieved data with the research community under an open-data license, creating barriers to open science. Relevant aspects include the amount of data being scraped \citep{klawonn.2019}, or the type of processing or aggregation being applied. To our knowledge, none of the articles reviewed for this work shared their data publicly.
\end{itemize}

Overall, research involving Web-scraped geographic data needs to address multiple, partly unique challenges at all stages from data collection and processing to data analysis and the interpretation of results. Nevertheless, using multiple digital data sources in concert may help to mitigate some of the biases inherent in each source \citep{tjaden.2021}.

\section{Sample Application: Leipzig Apartments}\label{casestudy}

To illustrate the potentials and pitfalls of Web scraping in a case study, apartment listings from Leipzig, Germany, will be examined. While possible applications in geographic research are manifold \citep{bonetti.et.al.2016, boeing.wadell.2017}, our original motivation for Web scraping real-estate data for Leipzig and other cities was to obtain a variety of datasets that could be used in teaching various geographic data science methods to students of geography, allowing students to relate general geographic knowledge to real-world case studies. Sample applications include hedonic price modeling using linear and nonlinear regression models, or black-box predictive modeling using random forest or boosting methods.

For the intended application it was necessary to retrieve samples consisting of at least several hundred apartment listings with complete information in key attributes. In addition to apartment rent and size, the year of construction and apartment condition (especially renovation) seemed necessary. Complete address information is also desirable in order to account for micro-geographic factors that would be calculated using GIS.

\subsection{Feasibility Assessment}\label{feasibility-assessment}

Two leading real-estate platforms were initially examined, Immowelt (\texttt{immowelt.de}) and ImmoScout24 (\texttt{immobilienscout24.de}), but the latter was excluded as its ToS disallow automated data capture, and technical restrictions have been in place since autumn 2020. Immowelt's terms contained no specific provisions related to Web scraping or the intended uses of the data, and neither did the \texttt{robots.txt} file, accessed multiple times during the scraping period, impose relevant restrictions. Privacy issues were not of concern as contact information, which very rarely referred to private landlords, was not to be extracted. Overall, considering these factors, the amount of data to be retrieved relative to the overall size of the platform's database, and the copyright regulations governing academic uses in Germany \citep{klawonn.2019}, it was deemed legally and ethically acceptable to harvest data from Immowelt.

The scraped platform is considered the second largest platform at the national level. At the time of writing, ImmoScout24, Immowelt and Ebay Kleinanzeigen (\texttt{ebay-kleinanzeigen.de}) held overlapping sets of 958, 651 and 513 Leipzig offers, respectively, with regional variation in market shares (e.g., Berlin: 2034, 366, and 1461 offers, respectively; Jena: 71, 76, 73). The scraped sample may therefore be biased as specific types of apartments may be procured through other platforms or directly through intermediaries such as real-estate agents or social media (undercoverage). This bias may change over time as the market share of platforms may vary due to mergers (e.g., Immowelt and Immonet in 2015; \citealp{handelsblatt.2015.merger}) or advertising campaigns \citep{horizont.2022.ebay}, posing challenges for trend assessments in inconsistent time series. Apartment offers also remain open for different amounts of time; while this clearly under-weights highly-sought apartments when using a cross-sectional design at a single time point, it should not affect samples gathered over an extended time period. \citet{beresewicz.2017} discusses representativeness issues in Web-scraped real-estate data analysis at depth from a statistical perspective.

Regarding the technical feasibility, the platform provides an easy-to-use (but not publicly documented) URL syntax that allows scrapers to access listings easily based on arguments such as object type, place name, or search radius. Actual user interaction such as scrolling or clicking are not required to access the HTML-based listings. A login was also not required. An inspection of HTML sources quickly showed that scraping would be easy to implement using basic programming tools. Although historical apartment data was not required for this study, it is worth noting that the Wayback Machine Web archive has not archived apartment offers from Immowelt.

A prototype was first implemented and tested and then deployed to routinely scrape real-estate listings. R's \texttt{rvest} package \citep{wickham.2021.rvest} was used along with \texttt{robotstxt} \citep{meissner.ren.2020.robotstxt} for scraping and information extraction. Retrieval was scheduled at night times and with generous delays (\textgreater10 seconds) between individual calls. Searches were done based on the city's name and using multiple sorting strategies to obtain a better coverage. All scripts were optimized to catch exceptions that may be related to server errors, internet connection problems, or unexpected webpage structure or data formats. The software is therefore able to ignore rare exceptions, but care must be taken to detect possible increases in the frequency of error messages, data gaps or inconsistencies. We limit our analysis to the calendar year 2021.

\subsection{Quality Considerations}\label{quality-considerations}

Overall, during 2021 there were 89 days on which no records were retrieved. These gaps mostly relate to an undetected technical failure during the summer holidays and a hardware damage in autumn; nevertheless, a peak in scraped apartments after the interruptions indicates that only a small fraction of the normal number of scraped apartments had been missed.

Overall, data on 9904 apartments in Leipzig was retrieved. Depending on the exclusion criteria used, between 6248 and 6505 apartment offers was available for subsequent hedonic modeling, as detailed below. Table \ref{tab:immotablemissing} shows statistics on missing data.

As for data quality, it is believed that information on `hard', price-determining facts is of generally high quality considering the landlord's possible liability for incorrect information. Nevertheless, a small fraction had implausible information such as an impossible construction year (one apartment) or price per square meter (two cases). The year of construction may in some cases represent the year of renovation as 67 renovated apartments were reportedly built after 2010.

Address information was usually complete with street name and house number. Nevertheless, some effort had to be put into processing address information in order to account for a reverted order of street name and postal code / town or to remove redundant and possibly confusing information from the address information (e.g., name of the neighborhood or unique characteristics such as `building B').

Address geocoding using the Nominatim Web service \citep{clemens.2015.nominatim} was successful (in the sense of returning a coordinate) for 94\% of the apartments. This compares favorably with the 71\% success rate achieved by \citet{bonetti.et.al.2016}. A systematic assessment of failed attempts has not been made, but it appears that 173 coordinates were clearly outside the city of Leipzig.

Euclidean distance to city center (town hall) was calculated as a simple proxy for access to services and retail. This variable was imputed based on the postal code (6\% of all apartments). (Additional imputation rules or regression models were implemented for some other variables that are not further used here, such as energy efficiency class: 57\% missing.)

\begin{table}

\caption{\label{tab:immotablemissing}Percentages of missing and implausible data for selected attributes in the 9904 Immowelt apartment offers from the city of Leipzig.}
\centering
\begin{tabular}[t]{l|c|c}
\hline
  & Missing & Implausible\\
\hline
Monthly (net) rent & 0.00 & \\
\hline
Apartment size & 0.00 & \\
\hline
Price per square meter & 0.00 & 0.02\\
\hline
Running costs & 0.00 & 0.01\\
\hline
Number of rooms & 0.12 & \\
\hline
Year of construction & 33.07 & 0.01\\
\hline
Energy efficiency class & 56.76 & \\
\hline
Postal code & 0.00 & 0.05\\
\hline
Geocoded coordinates & 5.57 & 1.75\\
\hline
\end{tabular}
\end{table}

\begin{figure}

{\centering \includegraphics[width=13cm]{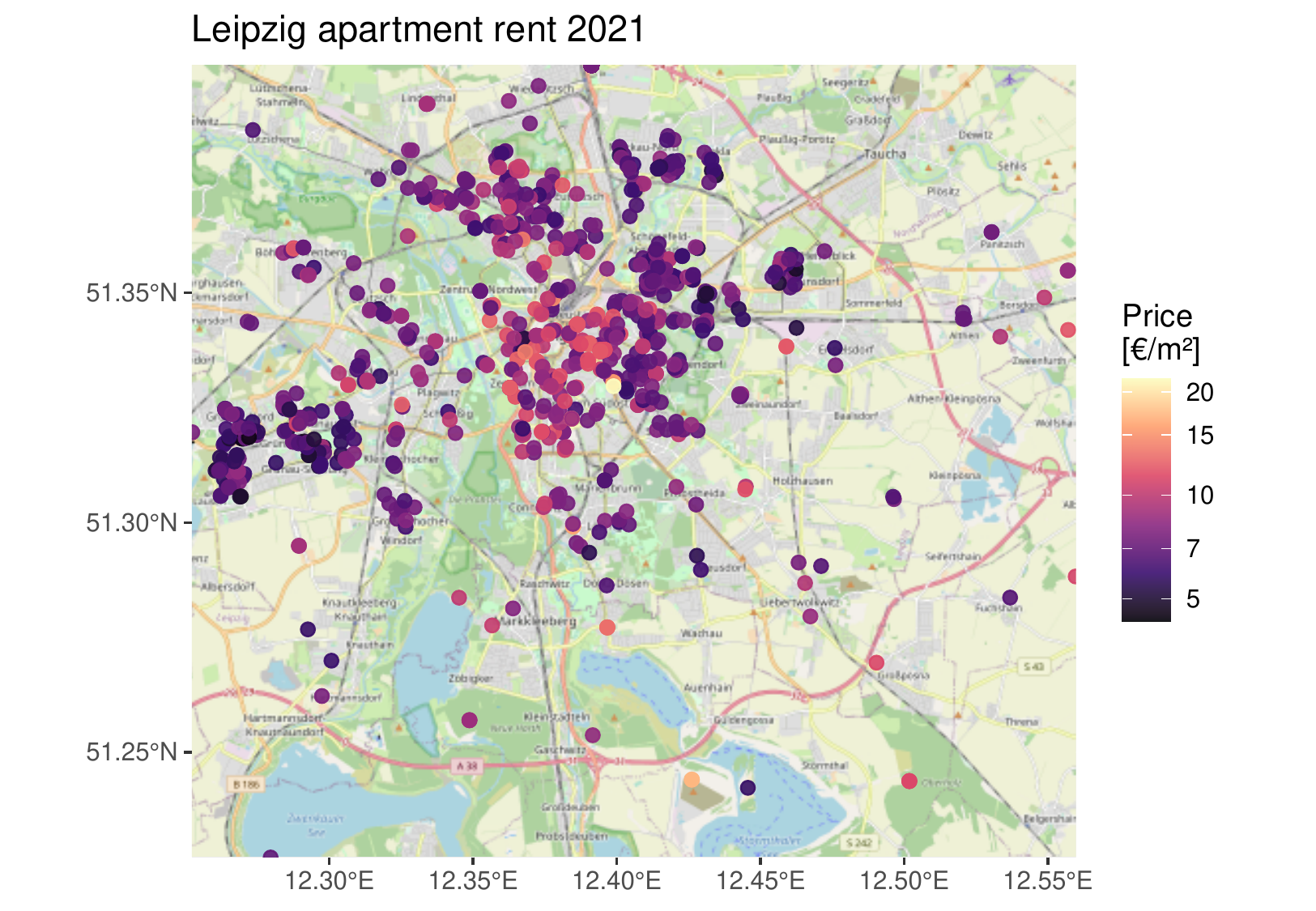}

}

\caption{Spatial distribution of a random subsample of 1000 apartments in the city of Leipzig used for training the models. Data source: Immowelt; map background: OpenStreetMap contributors.}\label{fig:immofigmap}
\end{figure}

Overall, substantially complete or imputed data on 6505 apartment offers from Leipzig were retrieved during 2021, with a more limited set of 6248 apartments having more precise building-level geolocation (Figure \ref{fig:immofigmap}).

\subsection{Geospatial Analysis}\label{geospatial-analysis}

In an academic teaching context, we use subsets of several hundred apartments for various activities, for example in practical assignments on geospatial data science, and to generate online quiz questions or exam questions using a programmatic literate programming framework \citep{zeileis.et.al.2014.rexams}. Random variability obtained by subsampling as well as varying sets of predictor variables (e.g., presence of a balcony, or fraction of green space in surroundings) offer a large amount of flexibility to obtain, e.g., significant or non-significant predictor effects or linear as well as nonlinear relationships. Additional variability can be achieved by varying the city, including cities with unique predictors such as distance to waterfront.

\begin{figure}

{\centering \includegraphics[width=15cm]{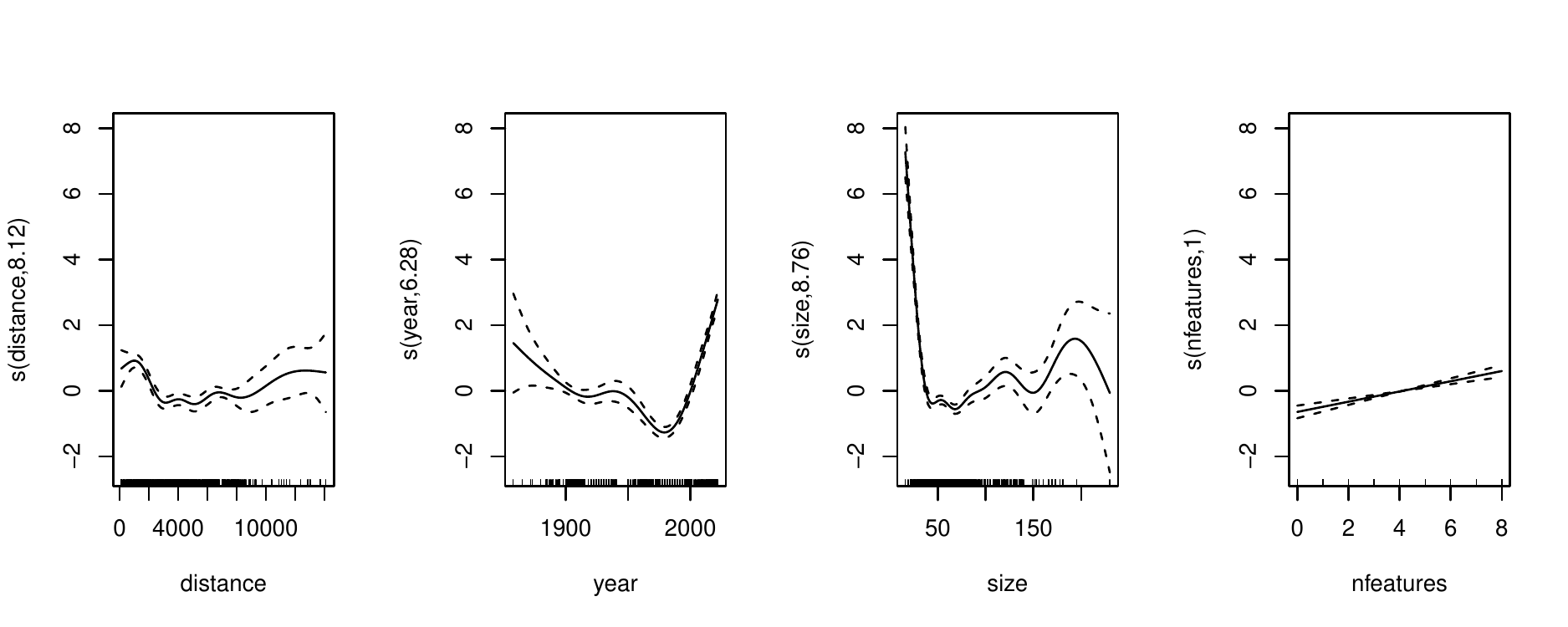}

}

\caption{Nonlinear transformation functions of a generalized additive model of apartment rent (price per square meter) in the city of Leipzig. Distance to city center is in meters, apartment size in square meters, year is the year built, and nfeatures represents the number of reported features such as balcony or parking space.}\label{fig:immofiggam}
\end{figure}

The example of a nonlinear regression analysis of apartment rents (per square meter) in Leipzig is shown in Figure \ref{fig:immofiggam}, using a training sample of 1000 offers. This generalized additive model \citep{wood.2017.book} achieved a very good model fit, considering its limited set of predictors (\(R^2_\textrm{adj}\): 0.678). General patterns relate to a sharp increase in rent per square meter for apartments built in recent years and for micro-apartments. Locations in close proximity (\textless3 km) to the city center are also more costly. In this simplified analysis, the number of features is a lumped quantity that simply counts the presence of features such as balcony, parking space, or senior-friendly condition. These features were positively, but much less strongly related to apartment rent, while accounting for year and distance to center.

Machine-learning models such as random forests \citep{breiman.2001} offer more flexibility and therefore possibly improved predictive performances, although at the cost of reduced model interpretability. The random forest model is also capable of handling strongly correlated predictors. This allows us to incorporate, for example, a larger number of variables representing specific apartment features (e.g., renovation status, built-in kitchen) or micro-geographic variables (e.g., distance to public transportation, postal code area). For comparison, the GAM was also applied in a data-driven manner by enabling variable selection by shrinkage to zero effective degrees of freedom \citep{wood.2017.book}. Model performance was assessed using all 5245 apartments that were not included in the training sample. Prediction maps were prepared for exemplary apartment characteristics held constant throughout the study area (55-m² two-room apartment built in 2000 with balcony, parking space, and basement).

\begin{figure}
\centering
\includegraphics{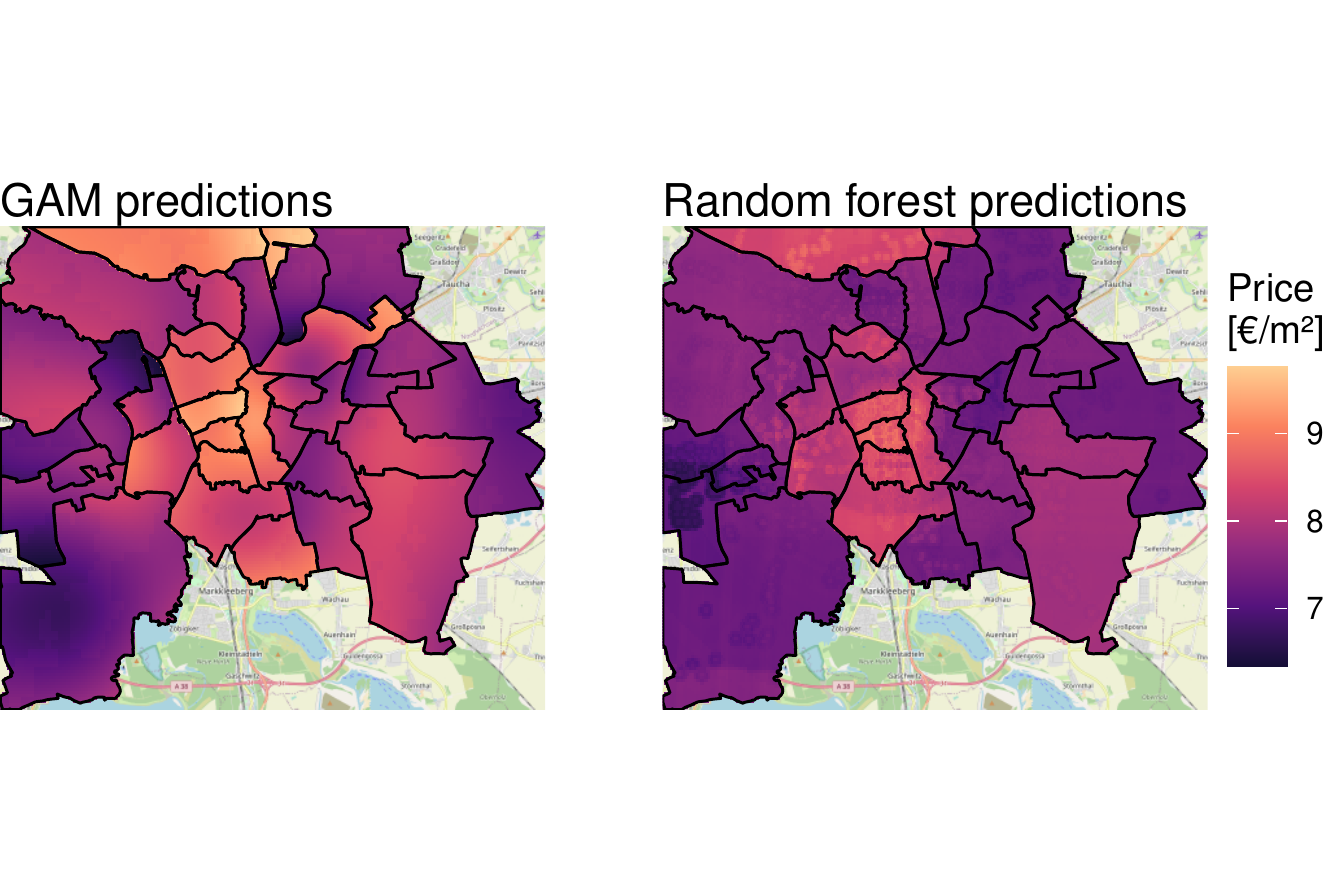}
\caption{\label{fig:immofigpredmap}GAM (left) and random forest (right) spatial prediction maps of expected apartment rent for a newly built two-room apartment with three features (balcony, basement, parking).}
\end{figure}

The random forest model with the enhanced feature set achieved the best performance on the test set (RMSE: 1.08 €/m²) followed by the GAM with shrinkage (RMSE: 1.22 €/m²) and the initial simple GAM (RMSE: 1.27 €/m²). Despite the similar performances, prediction maps in Figure \ref{fig:immofigpredmap} show important discrepancies that may relate to different model capabilities (smooth versus step-function relationships) and biases (variable selection bias of random forest: \citet{strobl.et.al.2008.conditional}).

\section{Discussion and Conclusions}\label{discussion}

The review and case study presented in this paper demonstrate that Web scraping as a tool for internet-based data collection and digital fieldwork offers numerous opportunities for geographic research, but also comes with several pitfalls. Considering the data openly available on the Web and the previously published research, Web scraping is particularly useful for the study of emergent patterns in urban transformation (e.g., gentrification, social inequality; \citealp{wachsmuth.weisler.2018.airbnb, mermet.2021.airbnb}) and platformed professionalization \citep{bosma.2022}, based on online platforms related to hospitality, real estate and mobility. Examples of studies exist in all fields of geography, making creative use of Web scraping to retrieve data on social and economic actors \citep{kinne.axenbeck.2020, gravelle.et.al.2021.mosques} or the physical environment \citep{canli.et.al.2018.landslides, skoulikaris.krestenitis.2020}) in order to understand spatial patterns and processes. Geocomputing tools such as geoparsing and geocoding as well as the enrichment of the captured data with ancillary location-based information show the relevance of geographic information science within the geographic Web scraping workflow.

Data that is purposefully provided for further analysis is often limited to privileged groups due to restrictions imposed by data owners or other gatekeepers. Access to and use of such data may result in costs (including the costs of paperwork; \citealp{vallone.et.al.2020}) or, if provided for free upon request, conflicts of interest as the permission may be revoked at any time. Web scraping therefore has the potential to democratize the access to data, at least within the legal, ethical and methodological limits laid out above, some of which have a geospatial dimension (e.g., geo-privacy, geoblocking, spatial undercoverage).

As an automated tool, Web scraping offers real-time insights especially into socio-economic processes that are otherwise monitored at annual to multi-annual frequencies through surveys and censuses, whose results can be delayed --- in some cases by years \citep{nyt.2022, zensus.2023}. This advantage has received increasing recognition by the statistics offices of many countries \citep{hoekstra.et.al.2012, virgillito.et.al.2019}, but its potential for creating prospective longitudinal samples for academic research is yet to be tapped. In some applications, Web archives may allow to extend time series into the past or to conduct retrospective studies \citep[e.g.,][]{tachibana.et.al.2021}. Nevertheless, the missing support for dynamic Web contents is still a strong limitation of he existing Web archives.

Although Web scraping is still partly uncharted terrain, researchers should not be afraid and should rather use the freedoms granted to them in most jurisdictions in order to generate datasets that would otherwise be difficult to acquire. Nevertheless, they should be aware that public access to contents does not automatically mean that the information is ``open'' or ``free'' in the sense of Open Science. We should therefore be cognizant of the legal and ethical limitations related to using and possibly sharing Web-scraped information. This is particularly true for potentially sensitive geospatial data.

%\bibliographystyle{dcu}
%  \bibliography{scraping.bib}

\end{document}